\documentclass[11pt,a4paper]{article}

\usepackage[T1]{fontenc}
\usepackage[utf8]{inputenc}
\usepackage{lmodern}
\usepackage{microtype}
\usepackage[a4paper,margin=1in]{geometry}
\usepackage{amsmath,amssymb}
\usepackage{hyperref}
\usepackage{graphicx}
\usepackage{subcaption}

\usepackage[round]{natbib}
\setcitestyle{aysep={}}

\title{The Good, the Bad, and the Ugly---Living with Priors in Bayesian confirmation}

\author{
Niels Linnemann \\
University of Geneva
\and
Christian Wüthrich \\
University of Geneva
}

\usepackage{subcaption}

\DeclareCaptionLabelFormat{figsub}{Fig.~\thefigure#2}
\date{PSA2026: The 30th Biennial Meeting of the Philosophy of Science Association (San Diego, CA; 19-22 Nov 2026)}

\begin{document}

\maketitle

\begin{abstract}\noindent
Bayesian confirmation faces a classic problem: where do initial priors come from? In cases with abundant data and repeated updating, different priors tend to converge to the same posterior. However, in frontier research this convergence often fails, and confirmation remains sensitive to priors. We examine how physics practice in the case of gravitational wave research deals with such cases and, normatively, when prior sensitivity should be regarded as epistemically problematic. We offer a practice-based account of the prior problem, drawn from a detailed case study in physics, of a kind so far  absent from the philosophical literature on Bayesianism. One upshot is a clearer diagnosis of so-called `analogue confirmation’.
\end{abstract}

\section{The prior problem and analogue confirmation}
\label{sec:intro}

Bayesianism comes with a wonderful feature: feed in enough data into your machinery, and you end up with effectively the same posterior, no matter your initial priors. However, prior insensitivity is, of course, only a special long-run phenomenon \citep{Earman}. 
In contrast, Bayesian inference in scientific practice often has to operate in circumstances of significant prior sensitivity (i.e., significant dependency on priors within the range of admissible priors). This sensitivity might be anything from mildly quantitative to sufficiently drastic to change the qualitative nature of the confirmatory context.

What to make of prior sensitivity? One may first of all wonder whether there is genuine freedom (and, if so, how much) in choosing priors. The issue of whether rational constraints can fix the prior has led to divisions within Bayesianism, with a firm `no' and a firm `yes' from extreme subjective and objective Bayesianism, respectively, and a `somehow' from subjective Bayesianism \citep{Titelbaum1, Titelbaum2}.
    
Accepting the current consensus that there is some freedom in choosing the priors, one may ask whether this is a bug or a feature. On a pessimistic note, it demotes Bayesianism into a conditional reasoning `calculus' on what is implied by certain subjective beliefs, rather than a normative account of confirmation. 
On a more positive note, this freedom is seen as allowing for the inclusion of background knowledge; on this understanding, Bayesianism is simply \textit{the} statistical approach that properly pays attention to our general state of knowledge, and not just to raw data. In fact, practitioners have long embraced the importance of priors for their set-up. 

In this paper, we provide a much-needed look at Bayesian confirmation from scientific practice to understand better how priors get set, and under which conditions they are seen to allow for confirmation despite prior sensitivity. Our case study will be that of gravitational wave observation, but we believe that our lessons generalise.\footnote{In fact, we have come across similar normative practices towards priors in many other fields. Take the field of clinical medicine where \cite{GoligherHarhay2022} note demands towards prior setting and treat them similarly to what we see in the GW practice: justification of priors, use of conservative neutral priors for primary tests, avoidance of standard priors, and sensitivity tests of priors. They also suggest checking adversarial skeptical priors.} 
Priors may be set by anything from plausibility considerations to a well established theoretical context. Overall, we will push back against the common conception that priors are just `guesses' in Bayesian practice. Among other things, we will draw attention to how priors are stated explicitly, how they are tested for sensitivity, and the drive with which scientists seek to replace them by more empirical ones.

One concrete application of our project clarifies the status of so-called `analogue' confirmation. Analogue confirmation is the idea that statements about a system can be confirmed by making observations on a different system that shares the same formal description. This idea has been particularly tempting in frontier research in black hole physics, where a phenomenon such as Hawking radiation is far from experimental reach, while it may be realised in analogous systems such as a Bose-Einstein condensate (BEC) \citep{stein2016} or a hydrodynamic tank \citep{rousseaux2008observation, rousseaux2010horizon, weinfurtner2013classical, euve2016observation}.
The question of whether analogue models can \emph{confirm} the existence of (gravitational) Hawking radiation in astrophysical black holes---more generally, whether analogue models can confirm hypotheses regarding inaccessible target systems---has been answered affirmatively by \cite{dareal17}, and criticised as circular by \cite{clw2021} (with \cite{field2025} arguably taking an intermediate position). Drawing on what we have learned about prior setting in practice (in our case study on gravitational wave detection), we show concretely how problematic it is that the Bayesian analogue confirmation framework developed in \cite{dareal19} exhibits high prior sensitivity. 

Our case study of Bayesian practice in gravitational wave research follows in \S\ref{sec:case}, with our analysis of the relevant lessons in \S\ref{sec:obs}. \S\ref{sec:analogue} applies these lessons to analogue confirmation. \S\ref{sec:conc} concludes.

\section{Setting the priors: the case of direct gravitational waves observation}
\label{sec:case}

For our interest in the actual practice of prior setting in Bayesian confirmation, the case of direct gravitational wave observation is particularly illuminating. It constitutes a highly non-trivial instance of observational confirmation at the frontier of modern physics, and its evidential status has significantly matured over time (since the first direct observation of gravitational waves in 2015, nearly 390 instances of gravitational waves in total have been reported at the time of writing \citep{GWTC5}), 
and, crucially for our purposes, it has largely been analysed by using Bayesian statistical methods.\footnote{Bayesian methods in physics, as in other sciences, have only been widely adopted within the last 30 years (one reason being computational costs, another perhaps a remaining frequentist `bias' in many researcher's statistical upbringings).} This case therefore offers an opportunity to investigate the impact of priors on confirmation as well as the \textit{perception} of scientists of the impact of priors on confirmation, and so to examine how confirmation may or may not arise depending on the status of the adopted priors.

Gravitational waves are perturbative wave solutions on a general-relativistic background. Examples of gravitational waves were predicted already by \cite{Einstein1916, Einstein1918}. However, theoretical consensus on their existence emerged only gradually from the time of the first GR conference at Chapel Hill in 1957 onwards (with even Einstein himself having called them into question at some point). It took until 1979 \citep{TaylorFowlerMcCulloch1979} for their first indirect detection (via observations of the orbital decay of the first discovered binary pulsar, which was attributed to gravitational waves), and until 2015---nearly 100 years after Einstein's original recognition---for their first `direct'\footnote{On the notion of `direct' vs.\ `indirect', see \cite{ElderDirect}.} detection by the LIGO--Virgo collaboration (GW150914\footnote{Gravitational wave observations are labeled by date of detection in the format `GWYYMMDD', with `YY' as in `20YY'.}) \citep{GW1}. These now \textit{observable} gravitational waves originate from merging events of compact objects, most commonly two black holes, but also from mixed black hole--neutron star systems (such as GW200105, \citet{abbott2021observation}) or two neutron stars (e.g.\ GW170817, \citet{abbott2017gw170817}).

With this as background, let us now depict the different stages of evidential support for gravitational wave observation and their origins in merging events since 2015---and the role prior setting played in it.

\subsection{Direct detection of the wave}

Despite the dominance of Bayesian techniques in the gravitational wave literature, the first direct detection (GW150914) \citep{GW1} was still reported using frequentist rather than Bayesian methods. One might speculate that this choice was made to appeal to more widely ingrained frequentist standards; and, importantly, in order to evade the issue of priors.

From the Bayesian perspective on statistics, there is no such thing as a prior-free analysis. However, frequentists are not thought to score an advantage in not setting priors---they merely leave certain assumptions implicit.\footnote{Consider e.g. the proposal by \cite{Ashton2019}. It makes a case for assessing detection---especially in the long run---via Bayesian analysis, thereby also accounting for learned astrophysical signals in the form of priors. Thus, making explicit the assumptions about astrophysical signals from the outset can be considered a virtue.} One could therefore speak of presuppositions that, although not rendered explicit as priors, function as what might be called \emph{tacit priors}. 

\cite{ElderDection} has discussed the role of what we just dubbed tacit priors in the direct detection claims. She is specifically interested in theory-ladenness and her principal worry is that the detection procedure, which works effectively via matched filtering of model-based predictions (i.e.\ GR wave-signal predictions) with detector data, already presupposes GR---the very theory that gravitational-wave observations are ultimately meant to test.\footnote{Notably, there is another search `pipeline' (of, however, significantly lower statistical power) that presupposes much less theory-laden modelling and for which this concern is thus much reduced \citep{ElderDection}.} According to Elder, these circularity worries are ultimately circumvented in that there are several associated tests that show that the data genuinely confirms GR (in the sense that, with different data, GR could in principle have been disconfirmed). For instance, in the \textit{IMR consistency test}, the posteriors for the mass and spin of the original merging objects are calculated from low-frequency data; from these, posteriors for the mass and spin of the post-merger object are calculated through numerical relativity. This result is then compared to the posteriors for the mass and spin of the post-merger object obtained directly from high-frequency data. Prima facie, more general gravitational theories (i.e., theories that deviate from GR in the strong-field regime) might be preferred over GR when using the high-frequency part, even if GR performs very well on the low-frequency portion; and so deviations might be expected. However, no such deviation is found; thus, there is a form of confirmation for GR vis-à-vis such competitors.

Elder’s discussion represents an interesting variation on the familiar theme that the prior may fully determine the posterior. If that were the case, the resulting `confirmation' would be altogether inconclusive. In what follows, we will be concerned with the question of the extent to which the posterior may legitimately depend on the prior for the posterior to amount to a robust result (for actual Bayesian priors, not for what we called tacit ones above)---a somewhat more subtle challenge to the conclusiveness of confirmation. Formally, the posteriors are always a function of the priors, the likelihoods, and the structure of the Bayesian network. The kind of dependence we are interested in here is the effective sensitivity of the posteriors under variations of the priors, rather than just this formal functional dependence.

\subsection{Indirect observation of the merging events}

The gravitational wave observations are at least as much about obtaining information on the merging events at the origin of the detected gravitational wave as they are about their mere detection. The merger event, assumed to result from the coalescence of two compact objects, is characterised by eight intrinsic parameters (the component masses and spin vectors) and seven extrinsic parameters (sky location, luminosity distance, binary inclination, polarization angle, coalescence time, and coalescence phase). One might add more sophisticated intrinsic parameters (such as on tidal deformability of the binary---think in analogy to the tidal effects between moon and earth); or more theoretical parameters about the model itself to distinguish GR from non-GR parameters (an easy example is graviton mass).

Such parameter estimation of the origin event is standardly done using Bayesian analysis, rather than a frequentist one. In early GW observation, the priors for these parameters were nearly completely chosen as ignorance priors. Prior sensitivity was, importantly, \textbf{(i) highlighted}\footnote{We number important stages of prior setting with small roman numerals.} as an issue.

Consequently, \textbf{(ii) systematic prior sensitivity analyses} were undertaken, such as that of \citet{Williamson2017} and, more comprehensively, \citet{Vitale2017}. Such analyses reveal that certain parameters, in particular that of ``the mass ratio and spins, have a smaller impact on the waveform and are therefore harder to measure. Measurements of these parameters are more directly affected by the chosen prior'' \citep[1]{Vitale2017}. In the course of such analysis, it is acknowledged both how important it is to recognise the effect of prior choice,\footnote{See e.g. \citet[1]{Vitale2017} who assert that ``[q]uantifying the effect of the prior choice is therefore a crucial step to make solid astrophysical statements using GW data.''} as well as how the requirement of robustness under prior choice ultimately simply calls for more data.\footnote{E.g. \citet[4]{Vitale2017} again: ``Data will be more informative for future loud events and, eventually, more and more physical conclusions will become robust with respect to the details of the prior choice.''}

In later GW observations, it was explicitly found that then available \textbf{(iii) population-based priors} gave corrected---and, in important cases, quite different---posteriors compared to those the original ignorance priors had given (as was indeed to be expected from the prior sensitivity analyses considered above). 

The technique of using the whole set of observations to inform single observations is called \emph{hierarchical Bayesian inference}. Schematically, for single events $i$, one first obtains $p(\theta_i | d_i)$ where $\theta_i$ is the parameter vector (say, with 15 entries) and $d_i$ the observational data for that event. Once one has a catalogue of events put together (i.e., a list of estimated parameters for each event),\footnote{Such (gravitational transient wave) catalogues (GWTCs) were continuously built for all major observational runs (O1; O2; O3a, O3b; O4a, O4b) in the LIGO-Virgo collaborations; see \cite*{GWTC1, GWTC2, GWTC21, GWTC3, GWTC3_population, GWTC4, GWTC5}; GWTC-1 included 11 events (from O1 and O2); GWTC-3 around 90 candidates (from O1, O2, O3a and O3b); GWTC-4.0 218 candidates after including O4a; and GWTC-5.0 around 390 candidates after including O4b.} 
one uses that knowledge about the catalogue to back-inform on the singular events as follows:
\begin{enumerate}
\item Conceive of the events as drawn from $n$ populations. One may test whether the assumption of exactly $n$ populations is adequate (e.g., via mixture models or outlier tests). Let us assume that there is just one population for simplicity.
\item Infer the population distribution for the parameters, i.e., $p(\theta | \Lambda)$,
where $\Lambda$ denotes the hyperparameters describing the population. (These are `hyperparameters' because they refer to a higher level of population compared to the parameters introduced before that refer to singular events.)
\item Use the population distribution as a prior when computing $p(\theta | d_{i'}, \Lambda)$ for novel events $i'$ and, if desired, for previously analysed ones (via posterior reweighting).
\end{enumerate}
Using this technique, one effectively uses population statistics to replace naïve priors on $\theta$ assumed in single-event analyses.

As is normal for Bayesian methods, there are of course still priors. However, they have now been moved up into the hyperparameters of the population model. Thus, while the direct priors for single-event estimation become more empirical under the use of a population-informed prior, some theoretical assumptions remain---namely in the choice of the hyperparameters and their prior distribution.

\begin{figure}[h]
    \centering

    \begin{subfigure}{0.48\linewidth}
        \centering
        \includegraphics[width=\linewidth]{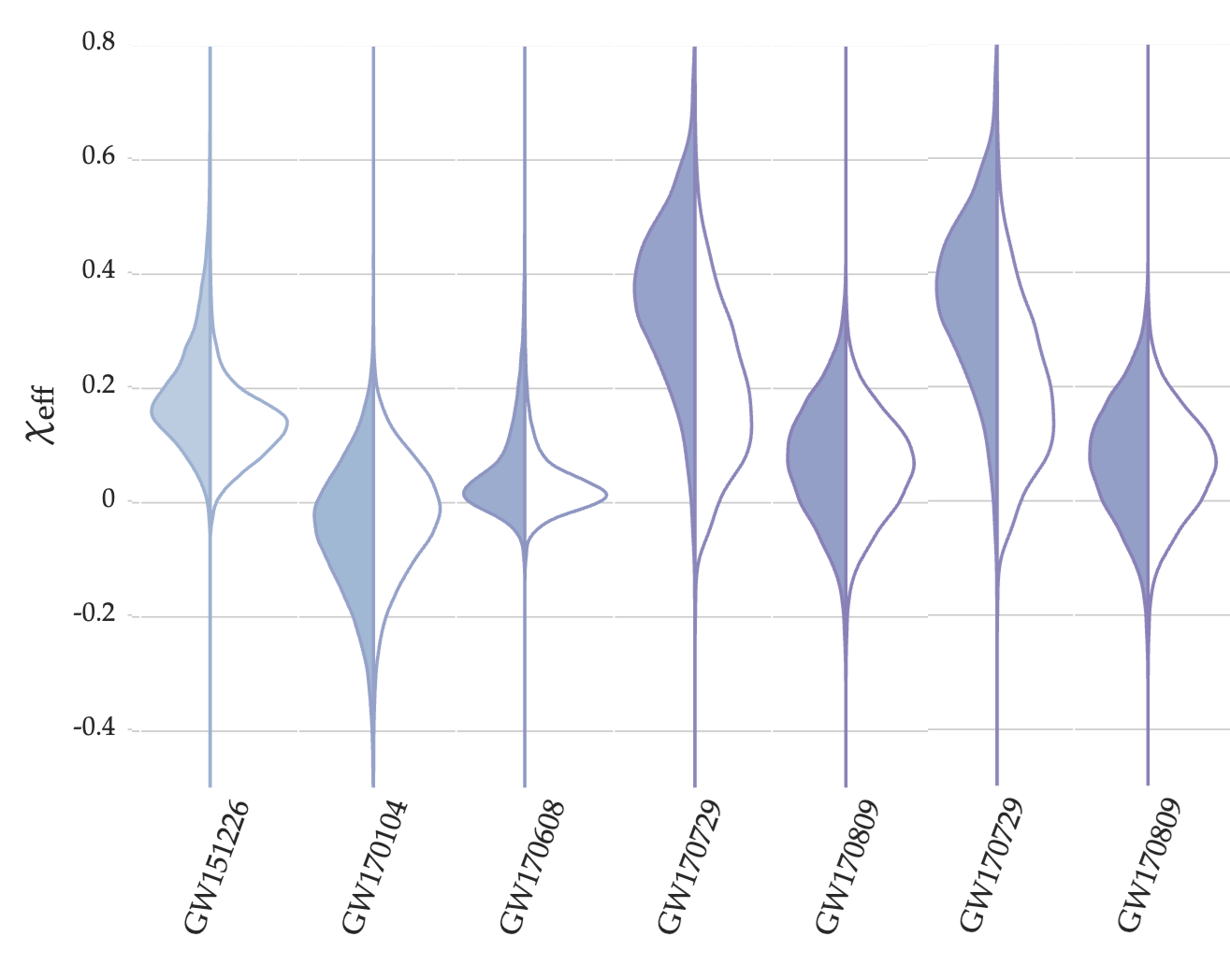}
        \caption{Violin plots of posteriors under default priors (filled in) vs. population-informed priors (white) for effective spin $\chi_{\mathrm{eff}}$ for various gravitational wave events, in order of detection. Adapted from \cite{MillerCallisterFarr2020}; original caption and the last three event entries removed for space.}
        \label{fig:effectivespin}
    \end{subfigure}
    \hfill
    \begin{subfigure}{0.48\linewidth}
        \centering
        \includegraphics[width=\linewidth]{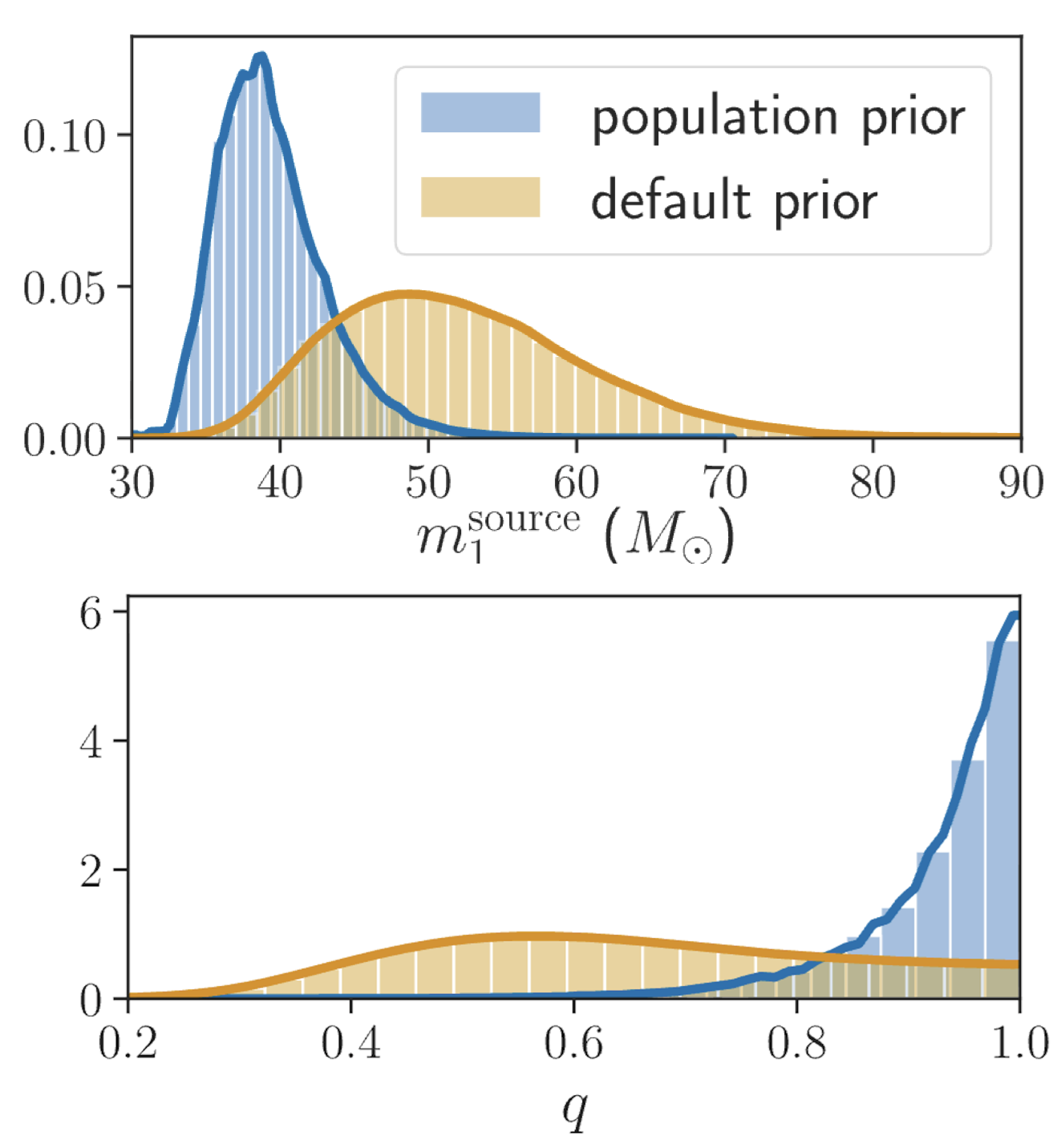}
        \caption{Comparison of default prior-based vs. population-prior-based posterior distributions (with different normalisations)  for primary mass $m_1$ and mass ratio $q$ of the merger event GW170729. Adapted from \cite{Fishbach2020}; original caption removed and panels rearranged.}
        \label{fig:masses}
    \end{subfigure}

    \label{fig:prior-comparison}
\end{figure}

Important corrections were thus achieved in particular for spin and mass estimates. \cite{MillerCallisterFarr2020} for instance note in their abstract that, after such analysis, ``GW170729, which previously excluded $\chi_{eff} = 0$, is now consistent with zero effective spin at $\sim$10\% credibility'', where $\chi_{eff}$ is the net projection of the spins of both merging components (black holes, or neutron stars) onto the binary’s orbital angular momentum. It is methodologically telling how they summarise their lesson: 
\begin{quote} A key takeaway from these results is that when a collection of binary black holes are analyzed in isolation under default priors common to such analyses, the magnitude of their effective spins will be consistently overestimated. When studying future compact binary mergers, it will be essential to evaluate their spins in the context of the broader population, particularly when an event seemingly has confidently positive or negative $\chi_{eff}$. (8-9)
\end{quote}
With respect to mass, one can also observe significant changes when hierarchical Bayesian inference is applied. For GW170729, for instance, the primary mass shifts from 51.2 to 38.9 solar masses ($\pm$ error) and tightens once a population-based prior is used \citep{Fishbach2020}. Figures \ref{fig:effectivespin} and \ref{fig:masses} illustrate graphically the change in the posteriors for spin and mass, respectively. As can be gleaned from the figures, the corrections for mass are even more dramatic than those for spin.

Another (albeit more limited) update on the priors for parameter estimation comes from observations of the same merging events through channels other than gravitational waves, known as \textbf{(iv) multi-messenger} astronomy. For GW170817, this was famously done for the first time: via optical observations one could identify the host galaxy and therefore determine the sky location and constrain the distance of the merger much more precisely. This information can then be incorporated when analyzing the gravitational-wave signal, effectively updating the priors on parameters such as sky position and distance in the gravitational-wave parameter estimation (see \cite{Properties2019} and references therein).

\section{The lesson from practice: confirmation is not confirmation}
\label{sec:obs}

Two core insights emerge from the case study of gravitational wave observation introduced in \S\ref{sec:case}.\footnote{Space constraints limit us to one case study here, but we take it that the charitable reader will follow us in how the lessons generalise. Everyone else can just treat this as an assumption on which our consequent analysis relies, and which we have hereby made explicit.}  First, at the early stages of scientific exploration of a phenomenon, the priors are set either as suggested by theoretical plausibility considerations or they are based simply on ignorance, often suggesting a flat, or nearly flat (a priori) prior distribution. Depending on the particular parameters of interest, the posteriors may or may not exhibit a significant sensitivity to how the priors are set. If they do, then scientists are wary of accepting the claim that a parameter $\theta_j$ lies in some set $S$ as \textit{confirmed}, even if the difference $\Delta:=P(\theta_j \in S|d) -P(\theta_j \in S)$ between posterior and prior is significantly larger than 0 and so in a flat-footed Bayesian analysis would qualify as confirmation. Rather than instances of confirmation, cases of prior sensitivity are taken to be prompts for further investigation. Although $\Delta$ may be large, the results then only have a preliminary status and a heuristic value. In this context, scientists take prior sensitivity to point to parameters for which improved empirical data is required before any robust confirmatory conclusions can be drawn. 

Second, the case study also suggests that once the priors are much more severely restricted by the rich statistics obtained from empirical data, even a small positive, indeed incremental, $\Delta$ may count as confirmatory. However, two central conditions for confirmation are, first, that scientists have a high degree of confidence in the priors and, second, that the posteriors are not too sensitive to the priors. In fact, these two conditions are not independent: the more sensitive the posteriors are to the priors, the higher the confidence in the priors needs to be; conversely, a lower level of confidence is acceptable in case the posteriors are not all that sensitive to the priors. To what degree the posteriors exhibit a sensitivity to the priors depends on the details of the Bayesian model in play. The confidence in the priors largely derives from the empirical handle the scientists have on them. Hence, this empirical handle needs to be the firmer the more sensitively the posteriors depend on the priors. 

Thus, we recognize that Bayesian analyses are used both at the early stages of exploratory discovery as well as later on in a much more rigidly circumscribed context of testing specific hypotheses. Comparing the two stages suggests that there is little correlation between the value of $\Delta$ and the degree of confirmation that a hypothesis commands---indeed, there may not even be any confirmation despite a large $\Delta$. Whether there is confirmation in any interesting sense depends on context---and in particular on the status ascribed to the priors used. So, $\Delta > 0$ turns out to be a necessary, rather than sufficient condition for confirmation.\footnote{Note that this also includes incremental confirmation. Pure incremental confirmation would apply to small positive---even epsilon-sized—deltas, but only if the priors can be regarded as reliable. Even for large deltas, we would not use the term incremental confirmation if the priors cannot be regarded as reliable. Otherwise, this would simply blur the distinction from the case of small deltas with secure priors. Sufficiently large deltas on their own seem to us---as reflected in the practice described---to only serve first of all as prompts for further investigation, that is, as assessments of plausibility and hence of worthiness of pursuit. Thanks to Stephan Hartmann for pressing us on this point.}

The recognition here that the scientific process happens in stages, and that the status of confirmation cannot be judged from a snapshot value of $\Delta$ without contextual information, arguably aligns with increasingly appreciated epistemological positions that stress the iterative or recursive nature of science.

First, it fits well with \citet{Chang}'s pragmatic take on the bootstrap between experiment and theory---or indeed the unfolding of science more generally---which he calls ``epistemic iteration''. 
The dynamics we observed for prior setting via population-based methods and multi-messenger approaches are exactly of this sort: results of one kind (single events of gravitational wave observation) are used to inform results of another kind (population estimates, or other observational techniques), which in turn feed back into the results of the first kind---and so on. Sometimes we get what Chang calls ``enrichment''; and sometimes self-correction. Just like on Chang's iterative coherentism, we do not start from a secure foundation but rebuild our ship as it remains afloat, permanently facing the danger that our efforts may fail---that the bootstrap might lead to ``self-destruction'' (\citet[ch.\ 5]{Chang}, who attributes it to \cite{Smith2002}).

Another, related, example is Smith's celebrated observation on how Newtonian gravity got substantially confirmed by what he calls ``closing the loop'' \citep{Smith}. According to this methodology, Newtonian gravity is not tested simply by whether or not it can account for the observational data. Rather, it is tested by whether it provides the internal resources to explain the discrepancies between an initial model and the data, and by whether the proposed explanations for those discrepancies---such as the postulation of an additional planet---are themselves confirmed by further observation. The parallel to our finding is that Smith does not methodologically accept a naïve confirmation statement (Newton gravity model gets confirmed by data or not; in our case, that would be to have a positive delta regarding a parameter $\theta_i$ to lie in some interval $[a, b]$) but sees the thrust of scientific efforts to be directed at using that result to create another, more subtle test case (producing new predictions from modelling the error; in our case, predicting data). Importantly, whether or not this succeeds is, once again, a matter of contingency.


\section{Application: analogue confirmation is not confirmation}
\label{sec:analogue}

The central idea of analogue confirmation is that the detection of analogue radiation in analogue systems could indeed confirm the existence of astrophysical Hawking radiation. 

\subsection{Bayesian analogue confirmation?}
\label{ssec:bayesac}

\begin{figure}[h!]
        \centering
        \includegraphics[width=0.4\linewidth]{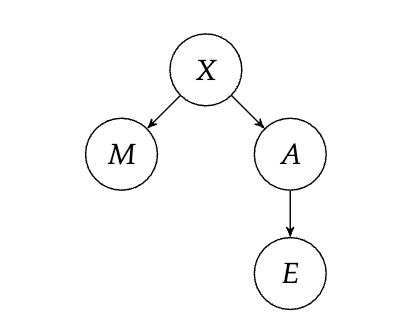}
        \caption{Bayesian network}
        \label{fig:network1}
    \end{figure}

\cite{dareal19} formalise this putative analogue confirmation as the Bayesian network of (binary) random variables in figure \ref{fig:network1} where the random variables stand for the truth of the following set of propositions, respectively: 
\begin{itemize}
    \item $M$: the semi-classical modelling framework provides an empirically adequate description of black holes, including their Hawking spectrum within its domain of applicability (DOA),
    \item $X$: ultra high-frequency physics does not affect the Hawking spectrum of systems in the relevant universality class,
    \item $A$: the semi-classical modelling framework provides an empirically adequate description of the analogue system within its DOA,
    \item $E$: analogue Hawking radiation is observed.
\end{itemize}

For this network, \cite{dareal19} prove a theorem showing that there is confirmation in the Bayesian sense of $\Delta := P(M|E) - P(M) > 0$ if certain conditions on the priors hold:
\begin{quote}
    \textbf{Theorem 1} Consider the Bayesian network with random variables $X, M, A, E$ in figure \ref{fig:network1} with a prior probability distribution $P$. Then $P(M|E) > P(M)$, if conditions $0 < P(X) < 1$, $P(M | X) > P(M | \bar{X})$, $P(A | X) > P(A | \bar{X})$, and $P(E | A) > P(E | \bar{A})$ are satisfied (where an overbar means negation).
\end{quote}
The status quo of the philosophical debate may be summarised, in the spirit of \citet{field2025}, by two exclusive options:
    \begin{enumerate}
        \item[(1)] The claim in analogue confirmation is of \emph{significant} confirmation.
        \item[(2)] The claim is of \emph{incremental} confirmation.
    \end{enumerate}

\cite{dareal17} arguably embrace (1) (talking of a ``high degree of warrant''), while \cite{dareal19} embrace only (2); in contrast \cite{clw2021} reject (1) and (2), while \cite{field2025} rejects (1)
, but embraces (2). In the rest of this section, we aim to show that both statements are problematic in light of our earlier case study.


\subsection{The decisive prior sensitivity}

The prior dependency of analogue confirmation is reflected in the formula:
\[
\Delta = \frac{P(X)P(\bar X)}{P(E)}\,\bigl(P(A|X)-P(A|\bar X)\bigr)\bigl(P(E|A)-P(E|\bar A)\bigr)\bigl(P(M|X)-P(M|\bar X)\bigr),
\]
where $P(E) = P(X)P(E|X) + P(\bar X)P(E|\bar X)$. 
Unless certain combinations of negative-valued differences are excluded, this means a decisive qualitative sensitivity of $\Delta$ on its priors. Given the practice in physics studied above, positive values of $\Delta$ should under such circumstances be interpreted not as confirmation, but rather as an incentive for further---empirical---research on ultra high frequency physics. 

Let us elaborate. If exactly one or three of the differences above are negative, we have in fact Bayesian \textit{dis}confirmation, i.e., $\Delta < 0$. Since it only makes sense to test analogue systems which fit the semi-classical framework and fall into the relevant universality class, the difference prior $P(A|X) - P(A|\bar{X})$ can safely be assumed to be positive. Since these analogue systems tend to be well understood theoretically and are empirically well confirmed, we can also assume $P(E|A) - P(E|\bar{A}) > 0$. 

So the difference of interest is $P(M|X) - P(M|\bar{X})=:d$. This also becomes apparent from the limit case of infinitely many different analogue systems probed (so an infinite sequence of evidence $E^{(1)}, E^{(2)}, ..., E^{(n)}, ...$ where each evidence is associated with a sufficiently different kind of analogue system\footnote{What this exactly means is worth discussing too but will be accepted for the sake of argument.}), \cite{dareal19} show that the confirmation becomes:
\[\lim_{n\rightarrow \infty} \Delta^{(n)} = P(\bar{X}) (P(M|X) - P(M|\bar{X}))=P(M|X)-P(M),\] where the last step is our (we hope instructive) rearrangement. (Arguably, if confirmation by $E, E',  ...$ goes through $X$, such a bound is not surprising.) The decisive prior sensitivity for any analogue confirmation claim of Hawking radiation in the framework of \cite{dareal19} lies exactly in that $\Delta$ (and its sign!) depend on the prior difference $d$. To repeat, for this to count as confirmation (even just of the incremental kind), our confidence in a positive $d$, and in its value, ought to be high. 

\subsection{Theory-informed priors for analogue confirmation}
\label{ssec:thyprior}

\cite{dareal19} set the priors solely per theoretical (plausibility) arguments, which seek to show that there is no so-called `trans-Planckian problem', i.e., that the ultra high frequency physics does not undermine the semi-classical model. Although they initially seem to be establishing some plausibility for the universality claim, the soundness of these heuristic arguments is easy to contest (see e.g. \cite{Gryb2021}). If you got a $\Delta > 0$ by setting the priors accordingly in light of such theoretical arguments, there would be no doubt for the practitioner---as discussed in our case study---that this is only an early stage result, pertinent for a context of discovery rather than a context of justification. From what we have learned, the prior sensitivity of the sign of $\Delta$---and thus of the confirmation claim---would be taken to show that simply more empirical study is called for.

To be sure, practitioners do use theory to conclusively \textit{exclude} priors based on well-established theory or on the very meaning of the parameters---but not on speculative plausibility arguments. In the case of analogue confirmation, an exclusion of this kind is perhaps available for $d < 0$, which would make the robustness of the Hawking spectrum count against the very description that predicts it, but not for the sceptical prior $d = 0$. \cite{clw2021} have argued that $d > 0$ is a mere presupposition: whether black holes fall into the relevant universality class---and hence whether an edge should be drawn between $X$ and $M$ at all---is the very research question at stake. \citet[\S4.3]{evansthebault2020} and \citet[26]{field2025} object that refusing a (strictly) positive difference amounts to inductive scepticism. But by \citet{evansthebault2020}'s own classification, the doubt is reasonable doubt after all; and, as the limit formula shows, no analogue experiment (no $E, E', ...$) can inform us on $P(M|X)$. 
Practitioners would usually require a confirmation claim to survive such a sceptical reference prior (for an explicit example from the medical sciences, see \citealt{GoligherHarhay2022}).
\footnote{You might also wonder that if a theoretical argument is to get you $P(M|X)-P(M)>0$, why then a theoretical argument could not deliver $P(M|E)-P(M)=\Delta > 0$, too, especially since the latter is bounded by the former.}

\subsection{Empirically informed priors for analogue confirmation?}
\label{ssec:empmicro}

Of course, our confidence in the priors could be boosted by an improved understanding of the relevant microphysics. If we had a fundamental theory, in this case of quantum gravity, that would both be empirically firmly established and would imply that black holes radiate like black bodies as predicted by Hawking, then we could effectively set our priors to 1. Thus, our confidence in the priors would ultimately be grounded in the empirical confirmation of the fundamental theory. This empirical confirmation could, of course, be of a very indirect nature and would not have to proceed directly via the detection of astrophysical Hawking radiation---otherwise the very raison d'être of analogue gravity would be undermined.

However, if $M$ would thus be confirmed through a well established theory $T$ of the relevant microphysics, we should properly take into account our background knowledge $T$, i.e., we should really consider $\Delta' = P(M | E, T) - P(M | T)$. Since $P(M | T)$ is effectively $1$ in this case, $\Delta'$ should be effectively zero. Thus, the attempted empirical justification of the priors (from background knowledge tied to $T$) does not allow for establishing either significant or even just incremental confirmation in the case where the priors are sufficiently empirically backed. Although Hawking radiation could be confirmed by empirically confirming a fundamental theory of which it results as emphasized by \citet{clw2021}, if we already have a firm grasp on black hole physics, analogue confirmation would effectively not even add incremental confirmation.

\subsection{Empirically informed universality arguments?}
\label{ssec:empuniv}

Thus, either the priors are grounded in speculative theory and do not inspire enough confidence to result in confirmation or they are based on empirically established theory and render analogue confirmation obsolete. We read \citet{field2025} as proposing another way in which there could be analogue confirmation.
In her view, if we only had enough empirical information to establish the \textit{universality arguments} (rather than directly the microphysics of black holes), then we could set $P(M|X) - P(M|\bar{X}) > 0$.

Although initially plausible, there is a problem with this proposal: how do we find a universality argument whose assumptions do not already amount to knowing the full micro theory? All universality arguments known so far are unsatisfactory, even on Field's own assessment. At this stage, without a positive proposal, all that Field has offered is the logical possibility that there may be a satisfactory universality argument whose assumptions we could somehow back up empirically. 

But is such an argument physically possible? In fact, Field herself is rather pessimistic and thinks that successful universality arguments could only be constructed if we are in either of two situations: if we are confident about the microphysics of our candidate systems for the universality class, or if we directly empirically test the macro-behaviour of these systems.\footnote{\citet{clw2021} identify these two possibilities.} In the latter case, there would be no need for the entire project of analogue gravity, which was premised on the supposition that the macro-behaviour of black holes is not sufficiently empirically accessible to detect astrophysical Hawking radiation directly. In the former case of high confidence in the microphysics and thus in black holes being adequately described by the modelling framework $M$, we are simply back in the situation described in \S\ref{ssec:empmicro}. 

\section{Conclusion}
\label{sec:conc}

Put a bit provocatively, the problem with Bayesian confirmation is not its priors; it is the dearth of practice-informed critics (and proponents). Consequently, it is that even discussions intended to feed back into scientific practice overlook well-established standards and distinctions from that practice that would help in assessing whether confirmation has been obtained. The case of analogue confirmation illustrates this particularly clearly: authors such as \cite{dareal17, dareal19} and \cite{field2025} aim to make claims about `confirmation', and even speak of `high degree of warrant' or `significant' confirmation. Yet their static modelling of confirmation does not track the sophistication and the multiple stages involved in prior setting and assessment---and thus for the delicate balance between the input delivered by the priors, the confidence researchers vest in them, and the confirmation the models may deliver.

In Bayesian modelling in scientific practice, a high sensitivity of the posteriors on priors---often observed in exploratory research---thus marks the need for a correspondingly high confidence in the values assigned to priors. This confidence is only gained from a theoretically or empirically well controlled context. We have argued that these conditions are not met in the case of a claimed analogue confirmation of astrophysical Hawking radiation by the observation of Hawking radiation in analogue models in the lab.

\section{Acknowledgements}

We thank Maya Fishbach and Simona J. Miller for the permission to use, respectively,  figure \ref{fig:masses} from \citep{Fishbach2020} and figure \ref{fig:effectivespin} from \citep{MillerCallisterFarr2020}. We thank Stephan Hartmann, Simona J. Miller, Ruward Mulder, and Dominic Ryder for detailed comments on a draft version.  We also thank Feraz Azhar and Stephan Hartmann for an earlier discussion on priors as well as audiences at the 2026 Dubrovnik conference and the Geneva Symmetry group research seminar. 

We acknowledge the support of the Swiss National Science Foundation as part of the project \textit{Philosophy Beyond Standard Physics} (105212\_207951) (N.L.) and of the ERC Advanced Grant \textit{EPIFRONT: At the Epistemic Frontiers of Physics} funded by the Swiss State Secretariat for Education, Research and Innovation (SERI) (C.W.).

\bibliographystyle{plainnat}
\bibliography{sample2.bib}

\end{document}